# Non-reciprocal emissivity, partial coherence, and amplification of internal energy from photon recycling when thermal radiation is sourced within matter

G.B. Smith, A.R. Gentle, M.D. Arnold, School of Mathematical and Physical Sciences, University of Technology Sydney; Broadway, NSW, Australia

Abstract

Photons excited into an internal ground-state mode at temperature T display well-defined partitioning of photons among variable transport properties such as photon phase, lifetime and distance travelled since creation. These property distributions have a mean, and a maximum which sets the distance from an interface photon creation can contribute to thermal radiation. All thermally excited photons at frequency f have the same phase velocity set by their standing wave's optical index n(f). Internal fluxes and internal energy then depend on T and the mode density set by n(f). All photons per mode that strike an interface and are obliquely emitted are refracted, so their exit intensities are irreversible except when very weak internal attenuation occurs. Attenuation index k(f) at very low T ~0 K is small, so universal reversibility is then approximate, but when T increases refraction of exiting photons varies as attenuation index k(f) rises. Total emission remains reversible after transitioning through a non-equilibrium state with no other heat inputs. In equilibrium the densities of excitations that create and annihilate photons are in balance with photon densities, and emissivity depends on n(f), k(f), T and internal incident direction θ*. Exit spectral intensities containing strong resonant intensities from pure water and crystalline silica are modelled, and match data accurately. Intrinsic resonances formed within liquids and compounds are due to photon modes hybridising with localized excitations, including molecular oscillations and the anharmonic component of lattice distortions. They explain the many resonant spectral intensities seen in remote sensing. Each hybrid oscillator is a photonic "virtual-bound-state" whose energy fluctuates between levels separated by hf. Other features addressed; radiance due to solid angle change at exit, anomalous refraction, thermal recycling of internally reflected photons, fluxes within multilayers, and enhanced internal heat flux from "phonon drag" by photon density gradients under an external temperature gradient.

1. **Introduction**
(i) *Alternate models; classical, semi-classical and quantum*
The models currently in wide use to predict spectral intensity profiles emerging from all heated matter, and their use to establish the total radiative cooling rates from a body in thermal equilibrium originated in Planck's resolution of the spectral intensity dilemma posed by emission from a cavity. An earlier prediction based on a classical thermodynamic treatment of total radiant power exiting a cavity by Stefan was not duplicated by Planck' initial photon model until he added a factor 2.0 to each intensity. Extensions to emission from condensed matter (1, 2) relied on various assumptions. Exit intensities were assumed to follow the energy conservation rule defined by a Kirchhoff emissivity acting on cavity emission, not from intensities created within a sample's volume but from quantum sources arrayed along its exit interface. Lambert(3) originally proposed this idea to explain the cosθ directional profile in intensity he observed exiting a hot metal ribbon at angle θ to the normal. Planck assumed Lambert's profile was universal. In the volume models we present here cosθ is usually present, as interior based intensities are refracted at exit to conserve momentum of each transmitted photon. Cavity emission is different being direct, so its intensities do not contain a cosθ factor. Had Planck used this he would not have needed to add to each intensity a factor of 2.0 to reproduce the correct classical EM total intensities derived by Stefan.

For emission from materials Planck's cavity emission plus the Kirchhoff rule emissivity retained this factor 2.0. We will show that hemispherical emittance $\varepsilon_H$ based on cavity intensities modified by the Kirchhoff rule is not correct as exit refraction means exit mode intensities subject to internal loss are irreversible. Weak exit intensities in the limit that sample T approaches absolute zero are almost reversible as each internal mode intensity has attenuation index k(f)~0. Photons carried by modes within band gaps in semiconductors also create almost reversible exit intensities. At finite T photons impacting an interface and exiting must transfer into the neighbouring ground-state mode that ensures momentum conservation. That entry mode becomes temperature dependent as T and k(f) rise so reversal of exit fluxes at finite T does not retrace the original internal intensity. A blackbody component in all thermal radiation was justified by the entropy requirements of the *classical* version of the 2nd Law of Thermodynamics. Quantum thermodynamics does allow single photon fluxes to be irreversible while maintaining reversibility of total exit radiant power $P_H(T)$, provided the original heat input rate dQ/dt is first removed. In equilibrium dQ/dt = $P_H(T)$ and temperature T is common to input and output, so both entropy fluxes match and the 2nd Law is obeyed.

The use of spectral absorptance A(θ,f) to define Kirchhoff emissivity $\varepsilon_K(\theta,f)$ was based on reversal of exit intensity I(θ,f,T)(4) at frequency f. This ignored exit refraction and neglected the thermal consequence of each A(θ,f)I(θ,f,T). The volume-based intensity models in this paper prove that oblique emissivity ε(θ,f) from all samples is non-reciprocal at finite T whenever each photon within a mode has a finite lifetime. Exit refracted photons are then irreversible when they transmit into a neighbouring ground-state mode whose direction ensures conservation of photon momentum. The Kirchhoff-Planck (KP) model for thermal radiation from solids and liquids unfortunately often gave approximate but always inexact agreement with precise exit intensity data at finite T for exit intensities out to moderate θ, which dominate exit intensities at higher θ. Our models show errors from the use of reciprocity hence $\varepsilon_K(\theta,f)$ at finite T become more obvious at oblique exit and as photon lifetimes get shorter at higher temperatures. Effects omitted that led erroneous KP model responses to get close to some data but without exact fits include (i) internal critical angles associated with refraction (ii) the decrease in correct ε(θ,f) as internal reflectance rises (iii) the unnecessary addition by Planck of his factor 2.0 (iv) the assumption that Stefan-Boltzmann σ is universal. Our internal models show cavity radiance factor $\sigma T^4$ is unique to a cavity while $\gamma T^4$ where $\gamma=\sigma/(\pi^4/15)=8\pi k^4/c^3h^3$ is a universal factor applicable to thermal radiation from all matter.

The usual approach to defining standing wave free modes inside matter is summarised in the Hamiltonian in equation (1). Each empty free photon stationary mode is occupied at finite T with spin σ photons, which are partitioned into energy levels $E_{k*}=hf=hc/\lambda$.

$$H_{ph}(k^*) = \sum_\sigma E_{k*,\sigma} c^+_{k*,\sigma} c_{k*,\sigma} = \sum_\sigma E_{k*,\sigma} n_{k*,\sigma} \qquad (1)$$

$k^* = 2\pi n(\lambda)/\lambda = 2\pi n(f)f/c$ is the internal wavevector with n(f) this mode's wave index at frequency f. All photons at T in such modes have phase velocity c/n(f) and $n_{k^*,\sigma}$ is the density of ground state modes with $k^*_\sigma$. Mode density $n_{k^*,\sigma}$ has both directional and volume components. A thermally created photon can propagate in any direction from its creation point so directional densities are high, spherically symmetric, and the same for all modes in the same material and different materials. Refraction means photons enter a neighbour's ground state mode that ensures momentum conservation. Volume density of modes then depends on n(f) for all bulk matter as defined in section 4. If other interfaces are close enough to an exit interface of interest extra internal mode impacts occur and modify mode density $n_{k^*,\sigma}$. At finite T the volume density of photons with $E_{k^*,\sigma}$ then depends on $n_{k^*,\sigma}$ and T. The dependence of $n_{k^*,\sigma}$ on $k^*$ means it as defined by n(f) as derived in section 4. Each free particle Hamiltonian is not simple as $n_{k^*,\sigma}$ now varies between sample modes and between materials but energy partitioning is not the only photon distribution at finite T that must be considered. Each created photon is subject to subsequent time and mode location-based partitioning. Equivalently all photons currently present have a well-defined "history" which can be defined by time-reversal as each created photon has taken a specific modal trajectory through time and space. The density of non-relativistic internal trajectories within a single mode obeys a distribution function which is derived in section 3.

Dual partitioning into energy then transport outcomes appears to be a central feature of quantum thermodynamics. Example properties into which photons at energy $E_{k^*,\sigma}$ can be partitioned include lifetimes $\tau(f)$, distances travelled $d^*(f)$ since creation at any mode location, and phase change experienced by each photon since its creation. Various balances in equilibrium are required to define the equilibrium distribution of photons present per mode among these properties. The required balances are listed in section 2 and include the balance between rates of creation and of annihilation. That balance ensures mode occupancy is stable and that the past experiences of all photons present at any instant hence the distribution density of photon phases present is also stable.

Each photon present can now be described by its energy and its current phase. The free photon ground-state Hamiltonian in eqn. (1) can be sub-divided as shown in $H_{k^*,\sigma,m}$ of equation(2). It defines the required dual ground state into which thermally generated photons will be dispersed, with m a transport property of interest.

$$H_{k^*,\sigma,m} = \sum_\sigma E_{k^*,\sigma} \sum_m^{n_{k^*,\sigma}} c^+_{k^*,\sigma;m} c_{k^*,\sigma;m} = \sum_\sigma E_{k^*,\sigma} \sum_m^{n_{k^*,\sigma}} m_{k^*,\sigma;m} \qquad (2)$$

$m_{k^*,\sigma;m}$ is the density of photons with energy hf within mode $k^*,\sigma$ with a specific transport property m available within each of its $n_{k^*,\sigma}$ modes, for example distance travelled $d^*(f)$ or phase $\phi(f)$. Our quantitative function describing partitioning among the m states within $n_{k^*,\sigma}$ provides a measure at each T of the partial coherence present in a mode. The rates that $c^+_{k^*,\sigma;m}$ and $c_{k^*,\sigma;m}$ operate must be in dynamic balance in equilibrium before we can define equilibrium intensities and the distribution of transport outcomes within each mode. The density at each f value of sub-states with property m obeys density $m_{k^*,\sigma;m}$ and at finite T measures the partial coherence among the photons within that mode.

Modes that strike an interface do not split into two. Their photons instead enter pre-existing modes, one in the initial material, the other in the material entered. The modes entered ensure that photon transfer conserves momentum. Those that tunnel to a mode in the next medium change their n(f), k(f) values so the mode they enter is defined by a complex Snell's law. Ground state modes can be defined using Maxwell EM waves when fluxes are not attenuated as T approaches 0 K. When T rises k(f) also rises so the neighbouring ground-state mode entered is temperature dependent as refraction depends on n and k, and unless k(f) ~0 exit mode intensities are irreversible. Optical phase changes following reflection by an interface do not modify phase distribution functions at finite T as each photon's phase shifts by the same amount.

A few models have addressed generation of thermal radiation inside matter. Their fundamentals differ to those in this paper, though there are common aspects. One used the semi-classical electromagnetic radiation from fluctuating electrical currents (5, 6) as defined by the fluctuation-dissipation (FDT) model's treatment of thermally excited quantum currents (7, 8). Another addressed an example of the many different spectral resonant features seen in remote sensing of minerals(9). The link of the Lorentz-Lorenz dispersion treatment of a classical oscillator to a material's complex refractive indices n(f), k(f) was used after general oscillator parameters had been fitted to observed emission spectra. The resulting n(f), k(f) for quartz matched known optical values. Their internal reflectance of obliquely incidence internal fluxes showed the Kirchhoff emissivity did not predict exit intensities though exit mode refraction was not addressed. Our models predict that special directional and spectral intensity characteristics occur after refraction of fluxes whose frequencies are near resonance. We include the role in equilibrium of photon thermal recycling following internal interface reflectance. Most alternate models do not address this issue, as they were surface based.

(ii) *Intrinsic resonant emission in thermal radiation*
Resonances within thermal radiation can be seen exiting most liquids, pure crystalline lattices, and matter containing local defects such as the vacancies and interstitials found in non-stoichiometric compounds such as $TiN_x$ and $Ti_yAl_{1-y}N_x$(10-12). These resonances occur when ground state internal normal modes can hybridise with a local oscillator mode. Photons then enter the hybrid for the resonance period, then return to the propagation mode. The supplement has further details on the mixing Hamiltonian involved and how a photon that enters then exits gains energy. In pure crystalline matter a high density of such hybrids per mode forms so their exit mode intensities are amplified and emitted intensities display strong resonant features. Two examples of strong resonant features our models predict precisely follow for (i) water where resonance occurs exactly at the known molecular mode frequencies (ii) a stoichiometric silica lattice. In crystalline matter harmonic distortions of each bond propagate as phonons, but are accompanied with localized anharmonic distortions that can hybridise with a photon mode of energy hf. Each local resonance occurs between energy levels $E_A$ and $(E_A+hf)$ with $E_A$ the energy of each anharmonic distortion which occur at high density in stoichiometric silica(9) and silicon carbide(13). The relative time a photon spends at energies $E_A$ and $(E_A+hf)$ relates to its energy gain at resonance. Intrinsic resonances within thermal radiation give precise spectral and chemical bond details, and occur inside liquids, solid dielectrics, many compound conductors and crystalline lattices.

(iii) *Photon transport statistics in each mode*

Section 3 covers the statistical distribution of equilibrium photon transport properties within states introduced in eqn. (2). As above they include distances $d*(f)$ travelled by each photon still present since its creation, and photon lifetimes $\tau(f)$. The distribution of these properties obey well defined rules based on equilibrium balances(14). They have mean values such as $<d*(f)>$ the mean-free-path which is a feature of distribution functions $P(d*(f))$ for each material. Equilibrium balance between creation and annihilation rates ensures that these distribution functions are stable and unique to each mode. They can be related directly to optical properties such as $\alpha(f)$ the optical absorption coefficient for each material. $P[d*(f)]$ provided us a proof that almost all emitted photons are created with a distance $d*(f) \sim 7.5 < d*(f) > \sim 7.5[1/\alpha(f)]$ from an exit interface, as $d*(f)$ spans values from near zero to a maximum $d*(f)_{max}$. $P(d*(f))$ ensures internal propagation usually occurs prior to emission and is derived and plotted in section 3. It allows us to set $d*(f)_{max}$ for each different material for a desired accuracy.

Photons and other thermal excitations such as phonons co-exist at finite T. The non-photon excitations that create and annihilate photons determine photon density in each mode and $<d*(f)>$(14). An addition to standard internal heat flow when an external temperature gradient is applied occurs as internal balance requires that a photon gradient matches an accompanying phonon gradient when loss and gain occur in photon-phonon collisions. "Phonon drag" by photons follows. Ground state mode directions within composite materials are initially uniform in each component material but mode topology changes if additional interfaces allow created photons added to a mode to have more than one chance of transfer into a neighbouring medium. Confusion also arises at finite T because the two ground state modes each photon can enter at an interface, which are another mode in its starting material or one in a different direction in the neighbouring material. These two choices result from momentum conservation hence a complex Snell's Law. The confusion results as *the mode entered within the neighbour is a function of temperature.* That is the new direction and intensity depend on n(f), k(f) in the two neighbours at T. Only in the external continuum is k(f) fixed at k(f) =0 as T rises.

The modelling principles we develop for linear modes are readily extended to some more complex internal structures, in particular multilayers, and select composites. Thermodynamic balance always requires the photon density along each mode whether linear or contoured to have well-defined $P(d*(f))$, $<d*(f)>$ and photon fluxes at each frequency. The rates at which photons enter from a neighbouring mode or leave to a the neighbour matter affect these responses. The local volume density of photons is no longer uniform if modes allow multiple escape chances into neighbouring modes along with the possibility of multiple internal reflections. Flux intensities and internal energy density matched when straight modes applied in all directions, but they become decoupled once the probabilities of multiple reflections are finite. Unlike classical many particle problems different local volume densities are allowed at finite T in quantum particle systems as balance involves thermal diffusion within modes not between them. Internal energy densities U(T) must then be worked out for each different composite structure. An example application of these principles to a structure with two internal interfaces and a note on what can be learned about internal mode structure generally from observation of external intensities I(θ,f,T) is in the supplement.

$P(d*(f))$ means quantum information is carried by thermally excited photons for use in various applications and that partial coherence per mode can be controlled and engineered in several ways. Quantum processes that drive the transition from a noisy, chaotic beginning to thermal equilibrium reduce noise content over time until balance sets in between creation and annihilation. Residual noise in equilibrium exit modes can also occur and will be demonstrated. Maxwell wave model treatments of thermal radiation, with FDT modifications added, have some common features to our quantum models. The FDT approach to photons is semi-classical and arose from the description of the response of thermally excited electrons subject to random ohmic loss within matter to applied fields. The resulting currents displayed Johnson noise(5-7). For FDT models of photon fluxes to reproduce the partial coherence and external thermodynamic equilibrium intensities predicted in this paper, and must incorporate exit refraction plus the thermodynamic balance rules following in the next section. A symbol glossary is included before our reference list as we are introducing concepts that may be unfamiliar to those used to standard thermal radiation models. The table also contains a useful summary of the physics used and introduced.

2. **Equilibrium intensities and precursor non-equilibrium states**

Non-equilibrium photon fluxes are irregular and precede the formation of the more regular transport properties in equilibrium. Photons flowing in transient and stable regimes have the same phase-velocity per mode, but transient densities are not in balance with the non-photon excitations with which they interact. The distribution of lifetimes per photon in each mode at finite T and energy hf influences the equilibrium spectral intensities within each mode and photon densities. Initial focus is on spectral intensities within bulk pure matter and the equilibrium intensities moving into the external mode that ensures each photon's momentum and energy are conserved. To exit photons tunnel through a terminating reaction potential into the required continuum mode. The Kirchhoff rule did not account for exit refraction despite mode index n(f) changing upon exit as all exit photons were created on the exit interface. Absorptance of reversed exit intensities to define emissivity also ignored outcomes from the additional heating rate δ(dQ/dt) it creates, with dQ/dt the heating rate that led to the equilibrium state at T=$T_0$ and output energy flux $P_H(T_0)$. Refraction at finite T always involves photons switching modes, which becomes increasingly irreversible as attenuation index k(f) grows and emissivity becomes increasingly non-reciprocal. The mathematical definition of oblique emissivity when k(f) is finite is derived rigorously in section 5 and never matches A(θ,λ). Refraction also traps internal propagating photons striking the interface above any internal critical angle $θ^*_C(f)$ due to total internal reflection (TIR), which adds to the internal thermal recycling of reflected photons whose $θ^* < θ^*_C(f)$.

Standing wave modes in samples with only one interface and sufficient size to be opaque, are linear and directionally spherically symmetric, but modes within multilayers and composites will display non-linear modes. A transmitting slab is the simplest example as its internal modes follow zig-zag paths between opposite interfaces as used by McMahon(15) and Kollyuk et al(16). Some features they predict our models duplicate by but basic errors occurred in both of these studies (see supplement), despite approximate agreement with select experimental intensities. One did not include loss from both sides (16) both neglected refraction and detailed balance within their zig-zag modes(15). Partial but inexact agreement with select optical data is common in older thermal radiation models despite their various errors which partially cancelled each other (see list above). Volume density of photons in equilibrium within extended linear modes is our starting point. Modifications for added interfaces depend on interface spacings. The supplement introduces this issue for multilayers as an additional statistical transport property arises, the probability a mode photon impacts an exit interface $w(f,\theta^*)$ times before it is finally annihilated or exits depends on interface spacing. Photon density along a contoured modes is constant in equilibrium, but volume density now depends on $w(f,\theta^*)$ as well as $n(f)$ and T and the length of each zig or zag segment between impacts.

Partial coherence in a mode emerges soon after initial heating, gets stronger, and finally stabilises. When a mode's annihilation rates rise phase coherence content grows, which is classically counter-intuitive but central to quantum thermodynamics. A related and possibly more puzzling corollary for classical physics is that as temperature increases photon internal annihilation rates rise as index $k(f)$ rises. The degree of photon coherence within a mode then grows. The primacy of Maxwell's models in the study of thermal radiation rather than photonic thermodynamics, was suggested by Mischenko(17, 18). That is not supported by our quantum thermodynamic models of modal fluxes whose stationary waves take account of all internal potentials that influence photon transport. Ground state solutions for nearly empty modes at T~0K when occupied at higher T should not be confused with the "rays" used in geometric optics. Each half of a standing wave solution at finite T carries matching but opposite energy fluxes. A Maxwell wave's amplitude and intensity reduces with distance travelled internally, as defined by absorption coefficient $\alpha(f)$, and is related to the survival probability of a photon with distance travelled from its random creation location in its mode. In thermal equilibrium at finite T, the following dynamic balances within and between fluxes are required. Predictions of individual flux intensities and equilibrium thermal outcomes all require the first three internal balances to be present. They are

(i) photon densities $N(f,T)$ and densities $M(f,T)$ of the excitations or defects present which create and destroy photons(14). The rates photons are created and annihilated both depend on $M(f,T)$ so they too are in balance.
(ii) the rate internal photons are recycled by internal reflectance at each interface and the rate these internally reflected photons are annihilated and add heat
(iii) the macroscopic rate of heat input $dQ/dt$ and total power radiated $P_H(T)$

Equilibrium balances (i), (ii), and (iii) apply within single materials and composites. An extra balance (iv) is needed for samples which have additional internal interfaces to the exit interface. They include transmitting slabs, coated substrates, multilayers, and matter containing particles or pores.

(iv) the rates that a mode's photons within each different material are lost by annihilation, by transmission to the next material including free space if applicable, and the rate all lost photons are replaced by creation events.

Resonances that result within samples with internal interfaces are structure based and distinct from the intrinsic anharmonic, bond defect and molecular resonances mentioned above. Outputs from some materials can inexactly approximate the spectral predictions of the Kirchhoff rule's predictions of resonant outputs from transmitting layers (15, 16) as their various errors partially cancel each other. The conclusion lists select historical examples where the Kirchhoff model could not be reconciled exactly with data, even when its predictions were close. Other balances to those just listed can occur, for example when photons are inducing motion in nearby particles or molecules.

Internal photothermal recycling in equilibrium can significantly impact all intensities. Due to the past focus on sources confined to an interface it was bypassed. Recycling followed by absorption resets equilibrium and internal energy density. Sample thermal mass C does not change but the thermal response to input dQ/dt does change, as internal heating rate is amplified. For a sample initially at ambient T=$T_A$, the expected stored heat change is $\Delta Q_0$ = C($T_0$-$T_A$)=C$\Delta T_0$ but annihilation of internally reflected photons adds heat so in balance total stored energy becomes $\Delta Q^*(T)$=C(T-$T_A$)=C$\Delta T$ with T>$T_0$ and $\Delta Q^*(T)$=$\Delta Q_0$/(1-$R_H$)= $\Delta Q(T_0)$/$\varepsilon_H$ (a proof is in the supplement). Internal hemispherical reflectance $R_H$ acts on internal hemispherical radiance $\Lambda_H$(f,T) Wm$^{-2}$Sr$^{-1}$ made up by all $\Lambda(\theta^*,f,T)$= $\Lambda(\theta^*,f,T)\cos\theta^*$ projected onto the interface. Thus $\Delta T/\Delta T_0$=1/$\varepsilon_H$ as internal heating rate is amplified. Heat generated by annihilation of recycled photons means equilibrium excitations M(f,T) and N(f,T) remain in balance but are amplified. Such changes must be accounted for in models of radiative cooling, and spectral and directional properties of external intensities. Total input and output entropy flows also change but remain in balance. In terms of externally applied heating rate dQ/dt a hybrid thermal heat capacity C*(T) determines equilibrium temperature T. Without recycling the usual result is $\Delta T_0$ = (dQ/dt)/C, but with recycling $\Delta T$ = (dQ/dt)/[(1-$R_H$)C]= $\Delta T_0$/$\varepsilon_H$ so hybrid heat capacity C*(T) = [(1-$R_H$)C] sets T. Since C*(T) = $\varepsilon_H$C(T) accurate $\varepsilon_H$ is important, whether from calorimetry, or directional emissivity. To the usual sensitivities governing C(T) we add for C*(T) the influence of $R_H$, any extra internal interfaces and exit surface topology. As $\varepsilon_H$ approaches its "white" limit of zero C*(T) also approaches zero and internal energy becomes quite large at fixed dQ/dt. Near the blackbody limit $\varepsilon_H \sim 1$, C*(f)~C(T) and neglect of internal recycling becomes a reasonable approximation.

The diverse sensitivity of exit intensities and directions to frequency are accurately predicted by our models. $\varepsilon_H$ governs radiative cooling rate $P_H$(T), but it is $\varepsilon_H\gamma T^4$ not $\varepsilon_{H,K}\sigma T^4$ from Kirchhoff-Planck models. $\gamma$=(15/$\pi^4$)$\sigma$=0.15399$\sigma$ with $\sigma$ the Stefan-Boltzmann constant. $P_H$(T) from accurate calorimetry does not change but $\varepsilon_H$=6.4939$\varepsilon_{H,K}$ is much larger than $\varepsilon_{H,K}$. The density of modes defined in section 4 using n(f) causes this rise. Starting with stationary ground state internal modes also avoids the need to use near fields induced by Maxwell waves at interfaces. The evanescent modes that traverse sub-wavelength pores and gaps between solid layers, or a small gap between a substrate and a nearby small particle or molecule are defined by the external extent of the reaction potential per mode at each interface. Whether fluxes emerge within internal gaps from these potentials depends on the size of gaps relative to the span of the potential which has the sharp form A(f)/$r^6$ as the Hamiltonian has an added "dipole-image dipole" reaction potential at the entry edge of any gap. The extent of this potential into a gap sets the probability an impacting photon is reflected, radiated across the gap, or transfers non-radiatively into the next material.

The distribution function P(d*(f)) is based on <d*(f)> from Fig. 1 next section. It defines the spread of photon phase changes since creation $\Delta\phi$(f) for each photon present in terms of mean phase change <$\Delta\phi$(f)>. The spread of all photon phases per mode at any instant also follows since creation and annihilation rates are in balance(14) The function P($\phi$(f)/<$\phi$(f)>) includes the phase of all photons making up photon density N(f,T).

P(ϕ(f)/<ϕ(f)>) and P(d*(f)/<d*(f)>) supply a variety of information including the value of $d^*(f)_{max}$, the partial coherence among photons within each mode and the proof that almost all emitted photons are created within a distance d*(f) ~7.5<d*(f)>~ 7.5[1/α(f)] from an exit interface. Both distributions are derived and plotted in the next section.

For a single material the magnitude of internal wavevector k*= 2πn(λ)/λ influences transmitted and reflected intensities at the exit interface. Its use to model a mode's contribution to volume density is described in section 4 which is focussed on large single material matter in equilibrium whose modes are linear and directionally spherically symmetric. Two useful optical identities that both prove that ε(θ,f) for a single material is non-reciprocal follow in section 5. One replaces the Kirchhoff model for ε(θ,f) given by $A_{TM}(\theta,f)$ or $A_{TE}(\theta,f)$ the spectral absorptances of reversed exit fluxes for each polarisation. We prove that ε(θ,f) = $|t_{TM}(\theta^*,f)|^2$ or $|t_{TE}(\theta^*,f)|^2$ for each internally incident intensity with t(θ*,f) the usual Fresnel transmittance coefficient. $|t(\theta^*,f)|^2$ never matches the Kirchhoff spectral absorptance A(θ,f) of a reversed exit flux except when θ*=0°. Reversal of oblique exit spectral intensities leads instead to internal intensities in directions different to θ* to the normal unless k(f)=0. The replacement emissivity acts on internally incident spectral intensities and the result accurately predicts the diverse spectral intensity characteristics exiting ionic and molecular materials. If IR indices n(f), k(f) are available no adjustable parameters are needed for accurate predictions.

## 3.0 Statistical transport properties defining partial coherence that also require internal propagation to precede emission

The distribution function P(d*(f)) models the survival probability with internal distance travelled d*(f) from the random creation location of all photons present within a transporting mode. A similar function was used to describe classical scattering loss from a beam of molecules with distance travelled by the beam(19). Quantum particle loss is different being a combination of photons not yet annihilated and those created to maintain equilibrium in place of those annihilated within a typical sampling time. That means at any one time in equilibrium a span of d*(f) per photon from 0 to $d^*(f)_{max}$ is set by a fixed distribution function P(d*(f)) of photon density with each d*(f). A formal derivation of mean-free-path <d*(f)> required for each P(d*(f)) is in equation (1), where <d*(f)> = 1/α(f). Internal intensity retains noise spikes, but they not usually detected in intensity data sampled over time periods of femtoseconds or longer for mid-IR radiation. A pictorial rendition of one set of creation events within $d^*(f)_{max}$ of an interface follows. It includes an example of exit noise when a photon is annihilated within $d^*(f)_{max}$ of an interface but not replaced by a compensating creation before its $d^*(f)_{max}$ cohort exits.

$$\langle d^*(f)\rangle = \int_0^\infty d(d^*)d^*(\exp(-\alpha(f)d^*)) = \frac{1}{\alpha(f)} = \frac{\lambda/2\pi}{\varepsilon_2(f)} = \frac{c}{\sigma(f)} = \frac{\delta(f)}{n(f)} \qquad (3)$$

The alternate optical parameters in eqn. (1) useful in defining <d*(f)> are $\varepsilon_2(f)$=2n(f)k(f), σ(f) the optical conductivity, and δ(f) the skin depth. The number of created photons present per unit volume that have survived transport over distance d*(f) is ΔN(d*(f)) = N(f,T)[exp(-α(f)d*(f)] while N(f,T)-ΔN(d*(f)) are the photons lost which on average are compensated by creation of N(f,T)[1-exp[-exp(α(f)d*(f))]] photons at random locations along the same mode to maintain equilibrium. The survival function ρ(d*(f))Δd*(f) =

$[dN(d^*(f))/d^*(f)]\Delta d^*(f)$ with $\rho(d^*(f) = n_{k^*,\sigma,d^*}$ of eqn. (2) after setting property $m=d^*(f)$. $n_{k^*,\sigma,d^*}$ with $f_{BE}(f,T)$ defines the number of photons present per unit volume that have been transported distances between $d^*(f)$ and $[d^*(f)+\Delta d^*(f)]$ since their creation. $\rho(d^*(f))\Delta d^*(f)/N(f,T)$ is then the probability a created photon will survive to between $d^*(f)$ and $[d^*(f)+\Delta d^*(f)]$. Replacing variable $d^*(f)$ with dimensionless variable $y(f)=d^*(f)/<d^*(f)>=\alpha(f)d^*(f)$ the probability distribution function $P(y(f))$ becomes upon evaluation of $dN(d^*(f))/d(d^*(f))$

$$P(y(f)) = -\{y(f)exp[-y(f)]\} \qquad (2)$$

which is plotted in Fig.1. $P(y(f)$ also guides the setting of $d^*(f)_{max}$ via a chosen pre-set $y(f)$ limit based on desired accuracy in setting the $d^*(f)$ cut-off beyond which a photon's transport is negligibly small. That means $d^*(f)_{max}$ also sets the maximum distance from an interface a photon can be created and have a chance of being emitted. $<d^*(f)>$, $P(d^*(f))$, and $d^*(f)_{max}$ are material specific properties that depend on annihilation rates. $P(y(f))$ means photons emitted are mainly created in the range $0.25<d^*(f)>$ to $2.5<d^*(f)>$ from an interface.

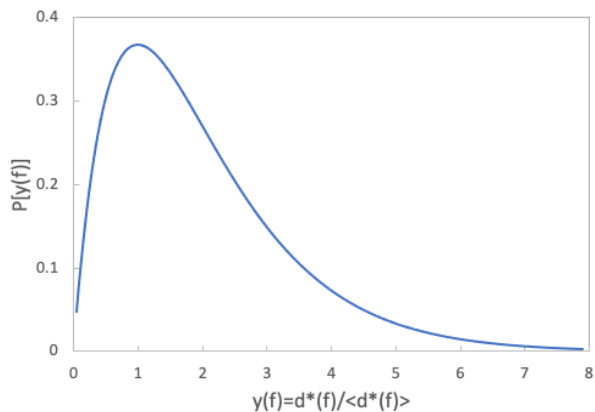

Figure 1. The distribution function within matter of distance travelled by each photon present from its creation point. Other photon internal transport properties subject to photon annihilation plus creation follow the same distribution function.

The function $P(d^*(f))$ also defines the partial coherence within each mode as it measures the extent of phase correlation between individual photons in a flux. Each photon present reaching $d^*(f)$ from creation has undergone a phase change $\Delta\phi(f) = 2\pi f[d^*(f)/c^*(f)]$ with phase velocity $c^*(f)=c/n(f)$. Averaging phase changes per photon over all photons present the mean-phase-change is $<(\Delta\phi(f)>=2\pi f\{n(f)<d^*(f)>/c\}$ or $(2\pi/\lambda)[n(\lambda)<d^*(\lambda)>/\alpha(\lambda)]$ in terms of wavelength. The distribution function $P(<(\Delta\phi(f)/<\Delta\phi(f)>)$ is $P(y(f))$ of fig. 1. Partial coherence within each absorbing mode follows if $<\Delta\phi(f)>$ is known as it also sets $P(d^*(f))$ and the photon phases in each mode. Coherence grows as the function $P(<(\phi(f)/<\phi(f)>)]$ narrows. The standout feature is that the smaller $<\phi(f)>$ or $<d^*(f)>$ becomes the narrower is the spread of the distribution of phase $\phi(f)$ per photon. That is *the phase correlation between photons in an internal mode is increased when the rate of photon annihilation increases*. This is counter-intuitive from a classical perspective where an increase in loss by scattering or friction is associated with less order and a rise in entropy flux. Increased loss by annihilation of quanta in equilibrium is different as it is compensated on average by matching rates of creation. The 2nd Law is not violated as thermal reversibility from reversal of hemispherical output occurs so input and output entropy fluxes match as $(dQ/dt)/T = P_H(T)/T$ match, with T set by internal recycling. Information and possible partial coherence within thermal emission have been previously noted(6, 8, 20, 21) but relied on Kirchhoff identities and surface sources.

As fig. 1 shows photons created beyond a transport distance $d^*(f)_{max} \sim 7.5 < d^*(f) > \sim 7.5/\alpha(f)$ from an exit interface have negligible chance of exiting. $P(d^*(f))$ and $d^*(f)_{max}$ are qualitatively distinctive if a material is a liquid, conductor, semiconductor, amorphous or crystalline. The schematic of photon creation sets shown in figure 2 is one of many sets of possible photon source emissions in one mode that contribute to internal and external intensity. Outputs are amplified each time a created photon enters then emerges from a hybrid resonant orbital and continues to propagate.

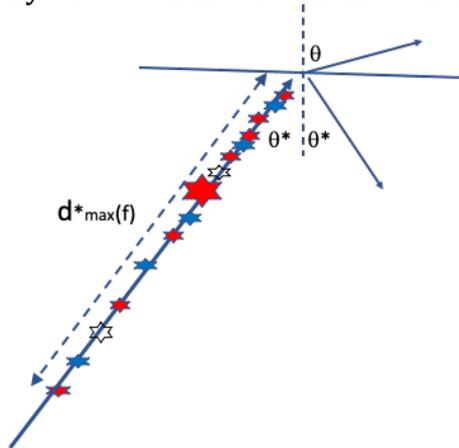

Figure 2. A schematic of one batch of photon creation events within $d^*(f)_{max}$ that exit. Coloured stars represent creation locations within $d^*(f)_{max}$ of an interface. Blue are photons replacing a prior annihilation. The empty star is a "photon hole" from annihilation not yet replaced and creates a fluctuation in exit intensity. The bright "star" represents a gain in mode energy after passing through a local hybrid mode. Final exit resonance is the sums all within several $d^*(f)_{max}$.

High exit intensities occur in limited frequency bands exiting crystalline matter such as SiC and $SiO_2$ as these modes contain a high linear density of orbitals resulting from hybrids formed with anharmonic lattice distortions which sum to the observed resonant intensity. An example section of a photon standing wave mode containing a regular dense array of hybrid orbitals formed with local anharmonic modes is sketched in the supplement. Lattice oscillations have two distinct components, thermally excited harmonic distortions which propagate as phonons and anharmonic distortions which remain localised within their originating bond. These distortions have other roles in thermal physics as they can scatter passing phonons(22). The resonances seen in thermal radiation from $SiO_2$ and SiC are between anharmonic energy levels $E_A$ and $(E_A+hf)$. Photons at energies hf enter the hybrid for $\sim h/(2\pi\Delta)$ seconds. $\Delta$ is the resonance width so extra local energy builds up which can be also be defined in terms of the phase shift from the time delay(23, 24). The local anharmonic energy for each lattice can be derived if desired. The supplement has a section containing a basic introduction to virtual-bound-states for this purpose. Previous VBS studies involved s-band free electrons in noble metals hybridising with localised d-orbitals on 3-d atomic impurities.

## 4. Internal spectral densities and intensities I(f,T)

Each internal ground-state mode has wavevector magnitude $k^* = 2\pi/\lambda^* = 2\pi n(\lambda)/\lambda$ with $n(\lambda)$ the real part of complex index. $\lambda^* < \lambda$ except inside metals or other materials at wavelengths where n(λ) < 1. If N(f) is the number of photon energy modes whose energy $hf_N$ < hf, for N =1 to N(f), the number of modes at energies between hf and h(f+Δf) is n*(hf)Δf = dN(f))/df)Δf with n*(hf) = 2n*$_{k*,\sigma}$ the mode density at energy hf from eqn. (1) for each spin. For large enough samples with no added interior interfaces n*(hf) is unique to that material. For samples with one or more interior interfaces the principles inside each large material apply except mode contours can change if multiple impacts with an interface are important. Equilibrium balances still apply throughout modes between interfaces, but they must account for photon loss to neighbouring matter at each interface impact, plus gain from mode impacts onto the other side of the same interface. Boundary loss rates from one side at each impact are identical, but different from matter on the opposite side. Thermal radiation displaying resonance features in coated systems where layer thicknesses allow multiple reflections per created photon have been reported(16, 25). Such structural resonances also occur in our equilibrium approach where mode contours and topology are determined at T=0 K so the exit direction is into a neighbouring ground-state modes at finite T already there. The localized hybrid resonances within stoichiometric crystalline matter thus have different origins to structure-based mode resonances between layers. The correct approach to accurate predictions of intensities exiting transmitting slabs is outlined in the supplement.

For different bulk materials with one interface accurate values of total internal energy U(f,T), uniform internal photon mode currents J(f,T) photons s$^{-1}$ and directional intensities I(f,T,k)=(hf)J(f,T,k) Wm$^{-2}$ for each internal mode, require a correct value of ρ(hf)=(hf)n*(hf). The expression N(f) based on the spherical symmetry of internal wavevectors **k*** defines the number of empty stationary states within an internal sphere of wavevector radius k* for each material. N(f) = (8πk*$^3$/3) after accounting for opposite spin photons in both half segments of two standing waves. Using optical index n(f)

$$N(f) = \frac{8\pi f^3 n(f)^3}{3c^3} = N_{BB}(f)n(f)^3 \tag{3}$$

The cavity standing wave density $N_{BB}$(f) = (8πf$^3$/3c$^3$) applies by itself only when index n(f)=1. Calculating (hf)[dN(f)/df] yields two separate terms. The second adds the term 3$N_{BB}$(f)n(f)$^2$hf(dn(f)/df) which we drop at this point as its contribution to photon density at finite T as shown in the supplement, is negligibly small relative to that from the remaining energy density term in equation (4). The energy level density ρ(hf)=n$_{k*,\sigma}$ from eqn. (1) with all internal fluxes in thermal equilibrium unpolarised is

$$\rho(hf) = 2(hf)n_{k*,\sigma} = \frac{8\pi hf^3 n(f)^3}{c^3} = \rho_{BB}(hf)n(f)^3 \tag{4}$$

and ρ$_{BB}$(hf)=(hf)[dN$_{BB}$(f)/df]. The internal intensity I(f,T) for bulk samples equals the contribution per mode to photon volume density N(f,T) = ρ(hf)f$_{BE}$(f,T) with f$_{BE}$(f,T) = [exp(hf/kT)-1]$^{-1}$ the Bose-Einstein occupancy factor. Converting the internal photon densities N(f,T) and intensity I(f,T) to functions of dimensionless energy units x(f,T) = hf/kT, yields internal spectral intensity

$$I(f,T) = I(x) = \frac{x^3 n(x)^3}{(\exp(x) - 1)}\gamma T^4 \quad Wm^{-2} \tag{5}$$

$\gamma = 8\pi k^4/c^3 h^3 = \sigma(15/\pi^4)$ with *0.15399$\sigma$ replacing the Stefan-Boltzmann constant for internally generated photons*. The expression for $I_{BB}(x)$ inside a cavity occurs only when $n(x)=1$ and as Planck found $I_{BB}(x)= \{x^3/[\exp(x)-1]\}\gamma T^4$. The photonic contribution to internal energy $U(f,T)$ within solid or liquid matter thus requires integrating of eqn. (5) over all x. When $n(x)=1$ the integration yields $U(f,T)=6.4939\gamma T^4 = \sigma T^4$. Within matter integration of eqn.(5) yields material specific multiples of $\gamma T^4$ which can be of order 6.5 when $n(x)$ is near 1.0 but are often much larger. The non-reciprocal emissivity we derive in section 5 adds an additional shift in spectral response. An example of the empty mode energy density $n_{k^*,\sigma}$ based on eqn. (4) is in figure 1(a) for silica and the resulting photon spectral density $N(f,T)$ calculated using eqn. (5) at 300 K is in figure 1(b). Figure (2) for water contains plots of $I(f,T) = (hf)N(f,T)$ at 300K and 360 K, plus normally emitted intensity $I(0,f,T)$ at 300 K as then $1-R(0^*,f) = 1-R(0,f)$. Internal propagation preceding emission thus adds important modifications to the Planck-Kirchhoff results in current use. Internally generated spectral intensities operated on by thermal emissivity are unique to each material. The second modification required is that oblique exit intensities have been refracted. For two element compounds with known complex indices in the IR, fluxes modelled provide convincing experimental validation that emissivity is internally defined and non-reciprocal.

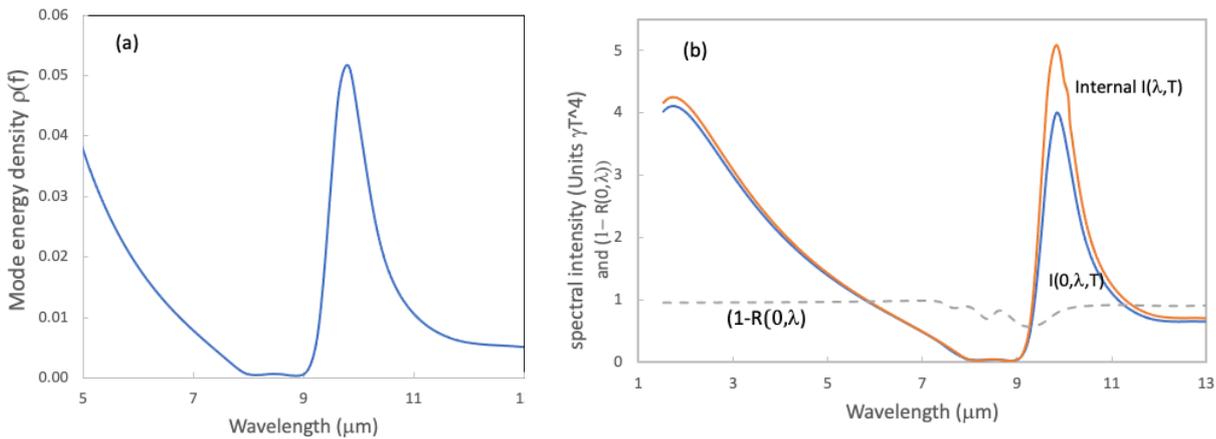

Figure 3. (a) Photon ground state mode density $\rho(hf)=hf\,^3n(f)^3$ inside silica in units of energy $8\pi h/c^3$ (b) Resulting internal spectral density $\rho(\lambda,T)$ at 300 K (red), and normal emitted intensity $I(0,\lambda,T)$ (blue). Intensity units are $8\pi k^4/(c^3h^3)]T^4$ and are universal. The blue plot in (b) is a close fit to observed normal intensities. Emissivity $\varepsilon(0^*,f) = (1-R(0,f))$ the dashed plot and is the only case where external reflectance can be used. Silica optical constants are from Kischkat et al(26).

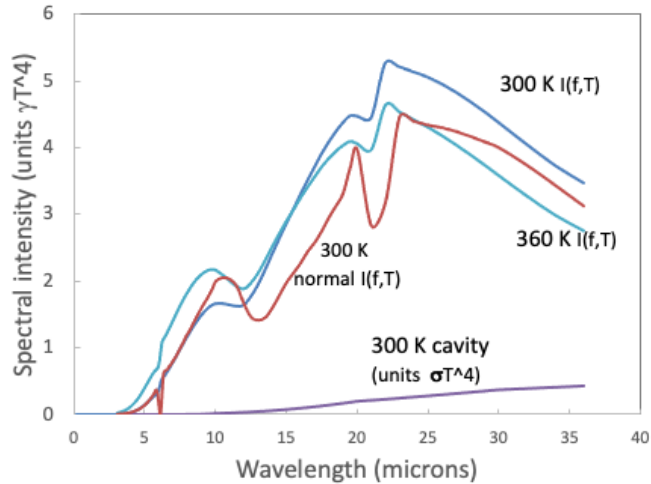

Figure 4. Spectral intensity at 300K and 360K within still water (blue) with the resulting normal exit intensity at 300K (red). Internal intensities in water display local photon resonances with water's four dominant molecular oscillation modes (27). Cavity intensities lack strong resonant features can be compared using the factor 6.4939. n(f), k(f) data used is from Hale et al (28).

Our predicted exit spectral intensities duplicate the variety of complex IR spectral intensities exiting compound materials seen in remote sensing. Overlapping internal resonances are commonplace and accurately predicted by these models as local resonances modify mode index n(f). Modes that form hybrids with photon mode in liquids are localised on molecules. Resonance linked to molecular mode oscillations within water are seen in fig. (3) using eqn. (5) plus correct emissivity. They are in excellent agreement with reported data from remote sensing studies(9, 29-31). Accuracy in predicted total and directional emission intensities, especially from water, are of growing environmental importance. Internal and external mode intensities in water display VBS resonances at its four main molecular vibration modes(27). These and the hybrid resonances in crystalline materials with anharmonic distortions means IR thermal intensity data and thermal image spectra can map materials within a mix. Both allow intensity data to be used for precision chemical analysis and the identification of molecular modes. Water's four resonant bands within thermal emission exemplify a common feature in thermal radiation from dielectric solids and liquids. They occur at energies that allow photon propagation modes and local oscillator or defect modes to add a local hybrid oscillator to ground state modes. Propagating photons "enter" each hybrid orbital for a brief period. After a short delay a photon reappears in the same basic mode at or very close to where it entered, with a jump in phase and energy to what it had upon entry. Photon and local oscillator momentum are conserved. Events like this in condensed matter physics are labelled "virtual-bound-states" (VBS)(23, 32, 33).

While photon internal density of states from eqn. (5) depends on $n(f)^3$ and T, internal mode reflectance, external transmittance and emissivity depend on n(f) and on k(f) the attenuating index. Refraction of the transmitted part of projected intensity **I**(f,T)cosθ* after projection of each I(f,T) onto an exit interface in direction θ* to the normal, defines emissivity. The reversal of the flux emerging in direction θ to the normal cannot reproduce the reversal of originating internal intensity **I**(θ*,f,T) as required by the Kirchhoff rule. Internal ground state fluxes are reversible at finite T if their k(f)=0, but once occupied most mode exit intensities are irreversible. Optical irreversibility of each different exit intensity I(θ,f,T) as noted already does not prevent thermodynamic reversal to the original equilibrium state if the *entire hemispherical radiant power $P_H(T)$ is reversed provided the original dQ/dt source is removed* (initially or over time).

Another important difference to emission from a cavity compared to that from matter after refraction is that the range of frequency modes exiting depends for matter on the frequency dependence of internal critical angles θ*$_C$(f). For some materials including plasmonic metals total internal reflection (TIR) is absent(34-36), but for many dielectrics θ*$_C$(f) plays a key role in emission. This is demonstrated in section 5, where our emissivity models are applied to refraction of radiation exiting bulk silica. Exit intensity directional profiles modelled for a few fixed θ* values vary widely as a function of mode frequency and but a reciprocal emissivity cannot predict such behaviour which will be observable. For dielectrics and for very hot matter refraction and TIR both vary emissivity. In some frequency bands exit intensity directions are anomalous as their exit direction θ < θ* for a limited range f after emissivity acts on internal projected intensity I(θ*,f,T)cosθ*.

When internal mode photons are subject to annihilation exit, refracted intensities cannot be reversed. The resulting non-reciprocal emissivity is formally derived from first principles using eqns. (1, 4, 5) in the next section. Spherical symmetry of all internal ground state modes means internal intensities I(f,T) are uniform in all internal directions. Ground state standing waves contain **J**(f,T,+**k**) and opposite flow **J**(f,T,-**k**), with half of each for up spin photons and half for down spin. This means Planck's addition of an extra factor 2.0 for exit fluxes was not justified, and would not have been needed had he not modulated cavity emission intensities with the factor "cosθ" as observed by Lambert(3) in his study of thermal emission from a heated metal ribbon. One of the emissivity replacement rules derived in the next section indicates that a "cosθ" factor always multiplies intensities exiting a smooth interface but it is due to the refraction required for *conservation of exiting photon momentum*. The one exception is intensity exiting a hole in a cavity wall as its internal fluxes exit without change of direction. For a contoured or rough surface, angle θ* is relative to the surface normal at each small area where an oblique incident photon tunnels through the interface reaction potential. In that case the "cosθ" rule applies locally. Planck's 2.0 is absent from all emission models when internal propagation precedes photon emission.

5. **Models for the non-reciprocal emissivity**

While photon internal mode density from eqn. (5) depends only on the real part of each material's complex index n(f), internal reflectance, external transmittance and emissivity depend on both n(f) and k(f) with k(f) a function of T hence of photon density in its mode. Exit intensity **I**(θ,f,T) flows within the external ground state mode in free space in direction θ to the local normal that ensures momentum conservation. It is the transmitted part of internal projected intensity I(θ*,f,T)=I(f,T)cosθ*. Whenever index k(f) within an occupied internal mode is finite, its value determines the direction of the neighbouring mode its emitted photons enter. The resulting expression for emissivity derived here is clearly non-reciprocal. Optical irreversibility of I(θ,f,T) does not however prevent optical and thermal reversibility of emitted hemispherical power $P_H(T)$ provided original heating rate dQ/dt is removed. Another difference to emission from a cavity is that the range of frequency modes exiting depends also for matter on interface structure and the frequency dependence of critical internal angles $\theta^*_c(f)$. For some materials, including plasmonic metals, total internal reflection is absent, but for many dielectrics it spans specific frequency bands for fixed internal directions $\theta^*$. We will demonstrate this by modelling select $\theta(f)$ after refraction of internally occupied modes exiting silica in section (6). A complex mix of exit directions results as frequency changes when $\theta^*$ is fixed. The Kirchhoff emissivity plus surface sources cannot predict such outcomes. In dielectrics and in very hot matter refraction not only involves TIR, but in some bands exit intensity directions are anomalous as $\theta(f) < \theta^*(f)$. This occurs as select exit intensities move closer to the normal than the internal angle of incidence which occurs either when $n(f) < k(f)$ (as in plasmonic conductors and dielectric Restrahlen bands), or when $n(f) > k(f)$ but they are close. Since $k(f)$ usually rises as temperature increases the prevalence of anomalous refraction is expected to increase with ongoing rise in T. An important example is exit intensities from stable plasmas as we expect them to cluster close to the normal to each plasma's boundaries. Our models in section 2 imply that at very high T a high degree of directional coherence in output fluxes is possible without invoking Zernicke(37) classical interference for a stable plasma in equilibrium but a rough plasma boundary is a non-equilibrium feature producing random emissions time and space. After travelling some distance into the continuum that spatial randomness may follow Zernicke's predictions and finally reproduce the expected spatial coherence.

If an externally applied thermal gradient $dT(x)/dx$ is present the steady state diffusion of heat and of photons can be based on a series of elemental equilibrium states where photon densities are $N(f,T(x))$ at $T(x) = [dT(x)/dx]\Delta x$. A photon density gradient $dN(f,T(x))/dx$ adds a photon diffusion current which does not transport heat, but local thermal equilibrium means that $(dN(f,T(x))/dx)\Delta x = [dM(f,T(x))/dx]\Delta x$. The gradient $[dM(f,T(x))/dx]$ is additional to that driving heat flux from diffusion of phonons in the absence of photons. Detailed balance within each $\Delta x$ segment sets local equilibrium. As a result photon density gradients $dN(f,T(x))/dx$ per mode "drag" phonons which add to the heat flow. Phonon drag of electron fluxes was postulated by Peierls(38) but his predictions of a $T^5$ dependence for low T electron currents was not observed. A contribution to heat flow from photon drag of phonons is expected in select materials as a function of temperature $T(x)$.

All exit intensities reduce to the form $I(\theta,f,T)=\varepsilon(\theta,f)I(f,T)\cos\theta$ which depends on the relation between $\theta$ and $\theta^*$ due to refraction. The small interface areas each hemispherical set of ground state photons converge to are of order a collision cross-section with the interface reaction potential, or about a bond length ($\sim 10^{-10}$ m) across. Emissivity acts on each projection $I(f,T)\cos\theta^*$. Two different identities for $I(\theta,f,T)$ arise and must produce the same exit intensity. First $I(\theta,f,T) = [1-R(\theta^*,f)]I(f,T)\cos\theta^*$ with $R(\theta^*,f)$ internal spectral reflectance, second $I(\theta,f,T)=T(\theta^*,\theta,f)I(f,T)\cos\theta^*$ with $T(\theta^*,\theta,f)$ the flux transmittance. Conservation of energy and momentum is then ensured by the identities in equation (6). The expression $I(\theta,f,T)= \varepsilon(\theta,f)I(f,T)\cos\theta$ for external intensity comes from the final expression in equation (6) operating on $I(f,T)\cos\theta^*$ so that $|t(\theta^*,f)|^2 = \varepsilon(\theta,f)$. Replacing $(1-R(\theta^*,f))$ with $\varepsilon(\theta^*,f)$ results in the alternate definition of emissivity in equation (7). The contribution of refraction to emissivity comes from the ratio $\cos\theta/\cos\theta^*$. A condensed conservation relation after refraction is $\varepsilon(\theta^*,f)\cos\theta^* = \varepsilon(\theta,f)\cos\theta$.

$$(1 - R(\theta^*, f)) = T(\theta^*, \theta, f) = |t(\theta^*, \theta, f)|^2 (cos\theta/cos\theta^*) \qquad (6)$$

$$I(\theta, f, T) = \varepsilon(\theta^*, f)I(f, T)cos\theta^* = \varepsilon(\theta, f)I(f, T)cos\theta \qquad (7)$$

Since ε(θ,f) = |t(θ*,f)|² defines emissivity with internal incidence angle used in a standard Fresnel amplitude coefficient it is straightforward to show that oblique ε(θ,f) never matches A(θ,f). Inclusion of the polarisation of internally reflected photons is required by equilibrium thermal balances and the two non-reciprocal emissivity expressions

$$\varepsilon_{TM}(\theta, f) = |t_{TM}(\theta, f)|^2 \; ; \; \varepsilon_{TE}(\theta, f) = |t_{TE}(\theta, f)|^2 \qquad (8)$$

Expressions for the Fresnel transmission and reflection coefficients for absorbing media are treated in many optical texts(39-41) and embedded in thin film software packages. θ* and complex indices n(f) and k(f), model each R(θ*,f) and ε(θ*,f). Observations of two or more exit intensities using eqns. (7) and (8) and the model for internal I(f,T) of eqn. (5) allows a material's complex indices at each wavelength to be extracted from a small number of observed intensities exiting in different directions. This is one validation test for this paper's models if the emitting sample's indices are known. Figures 3 and 4 were examples of successful theoretical and experimental validation tests.

To sum all external radiance elements to establish $P_H(T)$ the role of solid angle changes due to refraction must be included in radiance conservation rules. That is done in section 7. Steradian changes are essential to correct modelling of external radiance.

6. **The impacts of TIR and anomalous refraction on exit intensities**

To model spectral and directional outcomes defined by I(θ,f,T) expressions linking internal θ*(f) to externally observed θ(f) are needed to establish |t(θ*,f)|². There are two ways this can be done. A recently established approach is to use a complex Snell's Law (35, 36) recently derived for EM waves crossing the exit interface from within an absorber into another material or into the continuum. Expressions resulting from applying the usual interface boundary conditions are more complex than those in Born and Wolff(39) for external waves incident onto an absorber. The conservation rules from eqns. (7, 8) can also be used in place of complex Snell's Laws. If n(f) and k(f) are known all Fresnel coefficients needed can be evaluated then used to link θ*(f) and θ(f) since

$$cos\theta(f) = \frac{(1 - |r_{TE}(\theta^*, f)|^2)cos\theta^*(f)}{|t_{TE}(\theta^*, f)|^2} = \frac{(1 - |r_{TM}(\theta^*, f)|^2)cos\theta^*(f)}{|t_{TM}(\theta^*, f)|^2} \qquad (8)$$

If exit direction and exit intensity are known from experimental observations of I(θ,f,T) a solution or fit to θ* becomes possible.

It is instructive to study how θ(f) and I(θ,f,T) vary as a function of mode frequency for fixed θ*. Spectral exit intensities from each fixed θ* direction experience an interesting, and important array of directional outcomes from absorbing materials as frequency changes. Examples are plotted in figure 5 for silica of θ(f) from three fixed internal incident directions θ*. The frequencies chosen are for internal mode intensities at 300 K. Distinct frequency bands display the following sets of behaviour at fixed θ*; regular refraction, total internal reflection (TIR), and anomalous clustering of exit intensities close to the normal. The latter are labelled "anomalous refraction" since the exit direction is closer to the normal than incident direction $\theta^*$. Within dielectrics infra-red critical angles are often around 45° as seen in figure 5 for silica. Anomalous clustering near normal exit directions is also expected for plasmonic metals from recent studies of intensities exiting silver (34, 35). For metals and for dielectrics these refraction anomalies occur in spectral zones where $k(f) > n(f)$, or when $n(f) > k(f)$ but they are close. A large increase in $k(f)$ is expected within hot plasmas, so that its exit fluxes also cluster near the normal for an intense narrowly spaced set of intensities close to the normal to a stable plasma's edges. If the plasma is a sphere in equilibrium near normal directional coherence can explain for example the very small half-angle spread of solar intensities impacting our atmosphere despite the distance travelled to earth.

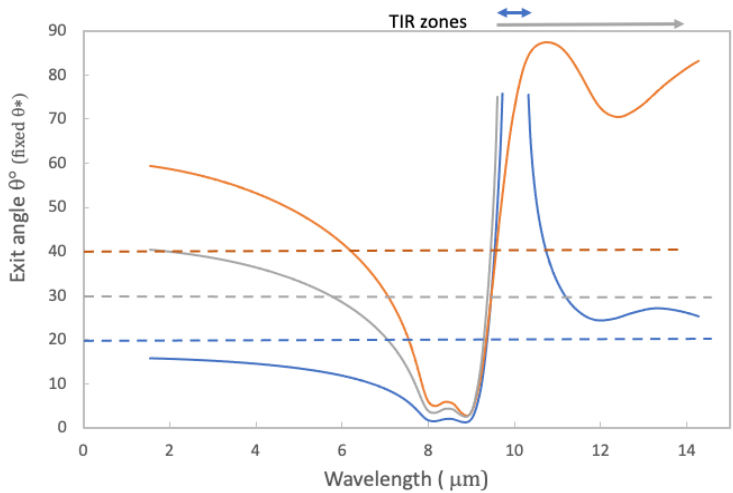

Figure 5. The spectral dependence of exit directions for three fixed internal impact directions $\theta^*$ (at the dashed lines) at 20°, 30° and 40° onto a smooth silica interface. The TIR band when $\theta^* = 20°$ (blue) is confined to the range 9 μm to 11 μm. In contrast for $\theta^* = 30°$ (grey) the TIR zone starts at ~ 9 μm but is not truncated by 15 μm. For $\theta^* > 46°$ in silica internal photons with $\lambda > 1$ μm experience TIR.

7. **Exit radiance and conclusion**

Exiting optical fluxes diverge so cooling rates requires consideration of the exit radiance after refraction resulting from internal radiance elements $\Delta\Lambda(f,T)$ Wm$^{-2}$Sr$^{-1}$. Like $I(f,T)$ internal radiance elements $\Delta\Lambda(f,T)$ are spherically symmetric in bulk matter at each frequency. Photons within modes making up $\Delta\Lambda(f,T)$ obey the same statistics and transport rules as single modes so internal solid angles $\Delta\Omega^*$ Sr are uniform and exit radiance has elements $\Delta\Lambda(\theta,\phi,f,T)=[\Delta I(\theta,f,T)/\Delta\Omega(\theta,\phi)]$ Wm$^{-2}$Sr$^{-1}$ after emissivity acts on the projection internally of $\Delta\Lambda(f,T)$ onto the interface for projected radiance $\Delta\Lambda(\theta^*,f,T) = \Delta\Lambda(f,T)\cos\theta^*$. The fixed internal sold angle $\Delta\Omega^*$ Sr means the internal spherical radiance element crossing through a small sphere is $(4\pi/\Delta\Omega^*)\Delta\Lambda(f,T)$ so $(2\pi/\Delta\Omega^*)\Delta\Lambda(f,T)$ Wm$^{-2}$Sr$^{-1}$ impacts the interface over many small areas. The energy-momentum conservation rule for emitted radiance then accounts for three geometric changes. They are internal to exit direction $\theta^*$ to $\theta$, internal to exit cross-sections dA* to

dA(θ,ϕ), and internal solid angles ΔΩ* to ΔΩ(θ,ϕ). The dependence of ΔΛ(θ,ϕ,f,T) on ΔΛ(θ*,f,T) derived as above for intensity to conserve energy and momentum is in equation (9). A dependence on θ* and ϕ* is initially included this time to allow for possible anisotropy in emissivity and internal reflectance when R(θ*,ϕ*,f) is dependent on ϕ*. With internal annihilation rates of photons required to balance internal reflection rates the radiance conservation rule now becomes

$$\frac{d\Lambda(\theta,\phi,f,T)}{d\Lambda(f,T)} = \varepsilon(\theta^*,\phi^*,f)\cos\theta^* \sin\theta^* d\theta^* d\phi^* = \varepsilon(\theta,\phi,f)\cos\theta \sin\theta\, d\theta d\phi \qquad (9)$$

For smooth surfaces dϕ* and dϕ in eqn. (9) cancel and the radiance conservation rule reduces to ε(θ*,f)cos θ*dΩ(θ*)=ε(θ,f)cosθdΩ(θ) for each polarisation with dΩ(θ) = sinθdθ, and dΩ(θ*) set at a constant internal but small steradian value.

The ratio [ΔΩ(θ,f)/ΔΩ*(f)] varies as exit direction θ changes due to refraction. It can be evaluated once the value of θ(f) has been determined for each internal radiance impact direction θ* as above. The extent of change in this ratio can be quite large and plays an important role in the exit radiance components making up hemispherical emittance $\varepsilon_H$, which total to cooling rate $P_H(T)$. Across anomalous refraction zones the ratio [ΔΩ(θ,f)/ΔΩ*] contracts near to the normal so that normal intensity can be intense. With axial symmetry eqn. (10) for $P_H(T)$ results with ε(θ,f) defined as before by eqns. (6-8) and dΛ(f,T) material specific replacing $d\Lambda_{BB}(f,T)$

$$P_H(T) = 2\pi\Lambda_H(T) = \int_0^{2\pi} d\Omega(\theta) \int_0^{\infty} df\, \varepsilon(\theta,f) d\Lambda(f,T)\cos\theta = \varepsilon_H \gamma T^4 \qquad (10)$$

This expression provides useful correction factors that can be applied to past data based on the Kirchhoff-Planck approach. These numerical factors can be close to 1.0, due to the various errors introduced by Planck which partially cancel each other. Errors can also be significant. Lambert's data was correct but his use of surface sources to explain it instead of refraction was not correct, and was unfortunately adopted by Planck and many others. A large amount of carefully acquired calorimetric emittance data since the 1950's to today could not be reproduced with optical calculations based on the Kirchhoff-Planck optical models. This paper's models can rectify that problem. A selection of such studies (42-46) are referenced. Many errors were attributed to unidentified experimental errors despite the careful analysis of possible error contributions that had been carried out. Their problem was instead the use of Kirchhoff-Planck intensity relations.

References and a Glossary summarising all main concepts

| Symbol/Term | Definition |
|---|---|
| n(f) | ground state standing wave phase index, real part of EM wave index |
| k(f) | photon dissipation index per mode, imaginary part of EM wave index |
| ε(θ,f) | spectral emissivity for fluxes in exit direction θ to local normal |
| ε(θ*,f) | [1-R(θ*,f)] at internal impact direction θ* to local normal |
| $n_{k^*,\sigma}$ | mode density at internal wavevector k* for spin σ photons |
| k*(f) | 2πn(λ)/λ, the internal wavevector of each ground state mode |
| $R_H$ | internal hemispherical reflectance |
| T | sample equilibrium temperature |
| $\gamma$ | universal constant, $8\pi k^4/c^3 h^3 = (15/\pi^4)\sigma = 0.15399\sigma$ |
| $\sigma$ | Stefan-Boltzmann constant (cavity emission only) |
| N(f,T) | internal density of photons per mode at T |
| M(f,T) | internal density of excitations that create and annihilate photons |
| dQ/dt | sample external heating rate |
| dQ*/dt | total internal heating rate due to photon thermal recycling |
| I(f,T) | internal intensity at f and T in all directions |
| I($\theta$*,f,T) | I(f,T)cos$\theta$*, Internal intensity projected onto the interface |
| $\Delta\Lambda$(f,T) | internal spectral radiance element at f and T in all directions |
| $\Delta\Lambda$($\theta$*,f,T) | $\Delta\Lambda$(f,T)cos$\theta$* Wm$^{-2}$Sr$^{-1}$ radiance projection onto an interface |
| $\Delta\Lambda$($\theta$,$\phi$,f,T) | external directional radiance at f and T |
| d*(f) | internal distance travelled by each photon since its creation |
| <d*(f)> | mean-free-path per mode (inverse of absorption coefficient $\alpha(f)^{-1}$) |
| d*(f)$_{max}$ | beyond d*(f)$_{max}$ photon survival probability is negligibly small |
| P[d*(f)] | photon survival probability with distance travelled from creation |
| y(f) | d*(f)/<d*(f)> |
| x(f,T) | hf/kT |
| $\rho$(d*(f)) | density of photons at d*(f) between d*(f) to [d*(f)+$\Delta$d*(f)] per mode |
| A($\theta$,f) | absorptance of reversed exit flux, also the Kirchhoff emissivity |
| $\varepsilon_H$ | reversible hemispherical emittance (based on irreversible I($\theta$,f,T)) |
| $\Delta\phi$(f), <$\Delta\phi$(f)> | phase change and mean-phase-change per photon since creation |
| C(T) | sample thermal mass without photon recycling by an interface |
| C*(T) | hybrid thermal mass based on dQ/dt with photon thermal recycling |
| $t_{TM}(\theta^*,f)$, $t_{TE}(\theta^*,f)$ | Fresnel complex transmittance coefficients for polarised emission |
| $\Delta\Omega$*, $\Delta\Omega(\theta,\phi)$ | common internal radiance solid angle element, refracted steradians |
| dA*, dA($\theta$,$\phi$) | common internal, exit directed external cross-sections per intensity |
| N(f) | $8\pi f^3 n(f)^3/3c^3$, the number of modes whose energy is below hf |
| $f_{BE}$(f,T) | 1/[exp(hf/kT)-1], the Bose-Einstein occupation factor per mode |
| U(T) | internal energy density from all excited quanta including photons |
| **J**(f,T,**k***), **J**(f,T,**k**) | internal and external photon flux vector "current densities" |
| dT(x)/dx | externally imposed thermal gradient in x-direction |
| TIR | total internal reflectance |
| VBS | virtual-bound-state between a photon mode and a local oscillator |
| $E_A$ | anharmonic oscillator energies that hybridise with photon modes |
| FDT | fluctuation-dissipation-theory (the semi-classical approach to the consequences of annihilation and creation of excited quanta) |

**Supplementary materials**

(i) Photon density and internal energy from photon recycling

The rate of energy feedback internally by an interface is based on equations (1) and (9) main text. Hemispherical emittance operates on hemispherical internal intensity $I_H(T)$ impacting an exit interface. The recycled radiant energy becomes $R_H I_H(T)$ with $R_H$ the hemispherical internal reflectance by the exit interface. The radiance conservation rules from the main text show that

$$\int_0^{\frac{\pi}{2}} d\Omega^*(\theta^*) \int_0^{\infty} dx \cos(\theta^*)[1 - R(\theta^*, x)] \frac{x^3 n(x)^3}{(\exp(x) - 1)} \gamma T^4 = (1 - R_H) I_H(T) \quad (S1)$$

with $(1-R_H)=\varepsilon_H$. Equation (S1) uses $R(\theta^*,f)=R(\theta^*,x)$ since $n(f)=n(x)$ and $k(f)=k(x)$ and the temperature dependence of $k(x)$ is needed in high temperature studies of $R(\theta^*,f)$. $I_H(T)=U(T)$ when internal fluxes are for samples with one exit interface whose dimensions are sufficiently large in all internal directions to span one or more $d^*(f)_{max}$ ranges. Then this $U(T)$ identity for internal energy density applies. It applies within each material when extra interfaces are present. Then $I_H(T)$ includes loss to and gain from neighbour matter in equilibrium. The integrand eqn. (S1) term $(1-R(\theta^*,f))\cos\theta^* d\Omega(\theta^*)$ before transformation to x can be replaced with $\varepsilon(\theta,f)\cos\theta d\Omega(\theta)$ as specified by eqn. (9) main text. This ensures preservation of energy and momentum so $P_H(T)$ the total emitted power by an interface of exit area A becomes

$$P_H(T) = A\varepsilon_H \gamma T^4 \quad W \quad (S2)$$

The product of each non-reciprocal emissivity with internal photon energy flux or radiance impacting an exit interface can be integrated to find $P_H(T)$.

Energy conservation requires that heating rate internally obeys detailed equilibrium balance (iii) listed in section 1 main text in which all internally reflected photons are being continuously annihilated in equilibrium. $dQ^*/dt$ is the internal heating rate due to the combination of external heat input rate $dQ/dt$ and rate of heat input produced by annihilation of internal recycled photons. Energy conservation requires the self-consistent feedback relation

$$\frac{dQ^*}{dt} = \frac{dQ}{dt} + R_H \frac{dQ^*}{dt} \quad (S3)$$

Solving for dQ*/dt gives

$$\frac{dQ^*}{dt} = \frac{1}{(1-R_H)}\frac{dQ}{dt} = \left\{\frac{1}{\varepsilon_H}\right\}\frac{dQ}{dt} \qquad (S4)$$

which mean the radiant cooling rate $P_H(T)$ required by detailed balance becomes $\varepsilon_H(dQ^*/dt)$. From eqn. (S4) this means

$$P_H(T) = \varepsilon_H \frac{dQ^*}{dt} = \frac{dQ}{dt} \qquad (S5)$$

which is the usual relation used in calorimetric studies. This cooling rate is different to that traditionally based on the bare heating rate $dQ/dt$ as the equilibrium temperature T can be much higher than the $T_0$ value expected without a contribution from photon recycling. The strong amplification factor $1/\varepsilon_H$ when $\varepsilon_H$ approaches its white body limit of zero was introduced in the main text. Low $\varepsilon_H$ requires $T \gg T_0$ hence an increased photon density, which means stored energy $U(T) \gg U(T_0)$.

(ii) Hybrid orbitals formed with molecular modes, local defects and anharmonic oscillations within a lattice

The four resonant bands for water shown in Fig.(4) and the strong single resonance band exiting silica in Fig. (3) main paper exemplify a common feature in thermal radiation from most dielectric solids and liquids. Resonance occurs when energies defining a photon propagation mode and the energy of a local oscillator mode overlap. A localized mixing potential is added at local oscillator sites to the basic Hamiltonian describing modes without defects in the main paper. Photon ground state modes whose energy coincides with one or many oscillator energies are modified by the addition of hybrid oscillators to each basic mode. Passing photons then enter each hybrid in its ground state mode where it picks up extra energy. If several hybrid bypasses occur in its mode the total resonant energy added in that mode is large. Emitted optical data indicates that two types of localized resonant modes occur

- those intrinsic to crystalline, defect free, stoichiometric compounds such as pure silica and silicon carbide. Their density of local resonance additions to ground state modes is high and repeatable for each pure compound, being based on anharmonic distortions which do not propagate

- those where select ground state modes form a hybrid oscillator at defect sites. Examples of these defects occur in thermal radiation exiting non-stoichiometric materials such as $TiN_x$(11) and $Ti_yAl_{1-y}N_x$(10) where defects can be interstitials, vacancies or ion substitutions at a lattice site such as Al at a Ti site in $Ti_yAl_{1-y}N_x$. The density of resonant modes in this case is defined by the stoichiometry hence the values of x and y along with the energy levels in the potential at the defect site

A schematic of a small section of a photon internal propagation mode where the local anharmonic distortion stores energy $E_A$ along with that of its partner harmonic component of bond distortion within an ordered lattice. The harmonic oscillations propagate as phonons while the excited anharmonic contributions remain in place and repeat at each stretched or distorted bond. Upon entering a local hybrid oscillator photon energy E=hf adds to anharmonic energy $E_{A1}$ to produce local boson energy $E_{A2} = E_{A1}+hf$. The hybrid then resonates between two local energy levels separated by $(E_{A2}- E_{A1}) = hf$. The relative probability of finding anharmonic levels $E_{A1}$ or $E_{A2}$ is then governed by the resonance width $\Delta(hf)$ using similar principles to that used by Anderson and Hubbard (32, 33) for virtual-bound states formed between free electron modes and defects formed by doping with atoms whose d-orbital energies match those of the host's s-band conduction electrons. A small section of a ground state mode containing hybrid photonic modes is shown schematically in Fig. (S1). A virtual-bound-state (VBS) resonance forms with the anharmonic component of distortion at each distorted bond.

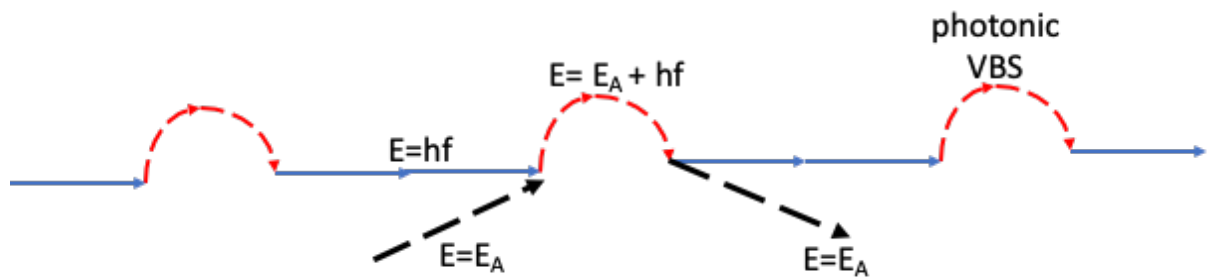

Figure S1. A schematic of a section of a ground state mode containing a high density of hybrid orbitals formed between a regular photon internal mode and an anharmonic oscillator. Mixing and de-mixing potentials $V_{k,EA}$ and its conjugate $V^*_{k, EA}$ operate at the junctions illustrated. The virtual bound state resonance width $\Delta$ formed applies between anharmonic energy levels $E_A$ and $(E_A+hf)$ separated by hf so hf and $\Delta$ together determine the relative probability of finding local energy $E_A$ or $(E_A+hf)$. The chemical analogue is a valence fluctuation.

The density of hybrid additions per photon ground state mode is high within stoichiometric crystalline lattices. They also raise propagation index n(f) slightly at resonance, but as a result internal photon density rises strongly through the impact of $n(f)^3$. The added energy when each hybrid orbital is occupied depends on the mixing potential strength of the coupling between the photon mode and local oscillator and its conjugate operator that separates the two modes (32) and allows the original photon to reappear in its free mode. This process can be thought of as equivalent to a mandatory "side-track" where photon transport is delayed while gaining energy before re-joining the main mode. A jump in the exit photon phase relative to what it would have had without the hybrid accompanies each side-track. Such local resonant events are well-known in condensed matter physics and were labelled "virtual-bound-states" (VBS)(23, 32, 33).

Localised thermally induced oscillations needed for resonance to occur are expected within liquid molecules, as we showed for water. Experimental data in silica exemplifies resonant hybrid modes formed inside stoichiometric crystalline matter. In silica local anharmonic oscillators hybridise with photon modes while the accompanying harmonic distortions propagate as phonons(22, 47). Lattice distortions thus play two distinct roles in photon internal transport (i) hybridisation of the anharmonic component of a bond's distortion (ii) generation of propagating phonons contribute to photon annihilation.

### (iii) Mode and photon density at finite T

Our photon spectral density models within matter that led to fitting of observed exit spectral intensity were based on the number of internal photon ground state modes $N(f)$ below frequency $f$. $N(f)$ was however limited to $(8\pi/c^3)f^2n(f)^3$ in final modelling though two terms arise mathematically upon evaluating mode energy density $\rho(f)=(hf)dN(f)/df$ which becomes $\rho(f)=(8\pi hf/c^3)[f^2n(f)^3+f^3n(f)^2(dn(f)/df)]$. For modelling experimental intensities at temperature T between $f$ and $(f+df)$ the conversion of $\rho(f,T)df = \rho(f)f_{BE}(f,T)df$ to a function of single variable $x= hf/kT$ is applied so that the energy density from $f$ to $(f+df)$ becomes $\rho(x)dx$. Upon doing this conversion the second term has one less $(kT/hf)$ factor than the conversion of the first term $[(8\pi h/c^3)f^3n(f)^3df]f_{BE}(f,T)$ which becomes $\rho(x)dx= [8\pi k^4/(c^3h^3)]T^4 (x^3n(x)^3dx)= \gamma T^4(x^3n(x)^3dx)$. The second term contains the factor $(dn(f)/df)df$ leaving $dn(f)$ and one less factor $(kT/h)dx$. The net contribution to output intensity of this second term has pre-factor $(8\pi k^3/c^3h^2)T^3$ which is lower by orders of magnitude than the pre-factor $(8\pi k^4/c^3h^3)T^4$. Neglect of this $T^3$ term does not affect accuracy.

### (iv) Composite matter with extra internal interfaces: equilibrium volume density of photons in each material

Optical models based on externally generated coherent wavefronts incident on a sample cannot be applied directly to intensities generated internally by heating. The outcomes of thermally excited photon creation must be described in statistical terms including the direction of the mode each photon enters, and their possible transport properties within that mode relative to that of all other photons in the same mode at the same temperature. Photon properties common to each photon within one directionally random mode within a dense spherical array are the phase velocity $c/n(f)$ and kinetic energy $hf$. Modes per material present where $T\sim 0$ where index $k(f)$ is close to zero do not change as T rises since $c/n(f)$ remains constant. The detailed paths traced out by a ground state mode no longer follow long straight lines. Our original thermodynamic restraints in equilibrium still apply to such modes though each photon created in them can now have multiple chances of exiting to a neighbouring material or to free space before it travels $d^*(f)_{max}$, beyond which it never survives inside the material in which it was created.

The number of possible impacts is variable as each photon created in modes between internal interfaces has a spread of its number of possible impacts with surrounding interfaces, governed by their spacing and the photon's mean-free-path $<d^*(f)>$. A transmitting slab of thickness L is the simplest case. Its ground-state empty modes can impact an interface few or many times and each photon entering one of those modes may strike an interface $w(f,\theta^*)$ times from $w(f,\theta^*)=0$ to $w(f,\theta^*)_{max}$, in integer steps which are material specific and depend on L. As before equilibrium requires that the rates at which photons strike each interface and are reflected or transmitted must be balanced by the rates at which those remaining are annihilated and add heat. Like survival distance $d^*(f)$, $w(f,\theta^*)$ is a variable transport property for photons created within each material in a mode set at angle of incidence $\theta^*$ to each layer. $w(f,\theta^*)_{max}$ can be expressed as $[d^*(f)_{max}/(2L/\cos\theta^*)]$. Our model for $P(d^*(f))$ is readily adapted to defining $P[w(f,\theta^*)]$ the probability an excited photon will experience w(f,θ*) reflections and transmissions. The mean number of internal reflections experienced is the integer below expression (S6)

$$\langle w(f, \theta^*) \rangle = \frac{\langle d^*(f) \rangle \cos\theta^*(f)}{2L} \qquad (S6)$$

Fabry-Perot resonances at different mode frequencies can occur within directional modes between parallel interfaces and are strongest when modes are normal to the exit interface(16).

## (v) Lambert's "cosθ "profile

The cosθ intensity profile originally observed by Lambert(3) from metal does not apply to radiation emerging from a small hole in a cavity wall. Planck assumed without proof Lambert's profile was characteristic of all thermal emission. His initial hemispherical power emitted from a cavity was then $0.5\sigma T^4$ or half the correct classical value, but without the cosθ factor he would have achieved the correct result while it presence for emission from matter is simply a consequence of internal propagation prior to emission.. A cosθ is a common intensity factor for emission from matter and adds support to our model and its non-reciprocal emissivity. Arrays of surface sources do yield a cosθ factor but ignore that both creation and A(θ,λ) hence annihilation occur internally along with the other thermodynamic consequences which we have shown require balance between creation and annihilation. The presence of TIR and bands of anomalous refraction within many materials add support to our approach. Each ratio $I_{TE}(\theta,f,T)/I(0,f,T)$ in the presence of refraction based on equations (3-6) main paper is

$$\frac{I_{TE}(\theta, f, T)}{I(0, f, T)} = \frac{|t_{TE}(\theta^*, f)|^2 \cos\theta}{(1 - R(0, f))} \qquad (S1)$$

with a similar result for TM intensities. When the two spectral intensities making up this ratio are separately summed over frequency at fixed θ, taking account that θ*(f) producing each experimental θ is sensitive to frequency, the ratio I(θ,T)/I(0,T) can be modelled but each net intensity at θ, will involve a range of frequencies hence of θ*(f) that satisfy eqn. (6) main text. Experimentally intensities are often proportional to cosθ but to achieve Lambert's empirical result of I(θ,T)/I(0,T) = Kcosθ, with K a constant is not guaranteed nor universal.

For supplement reference numbers refer to the main paper's reference list.